\documentclass[11pt,aps,prd,reprint,amsmath,superscriptaddress,showpacs,showkeys]{revtex4-1}

\usepackage{graphicx}			% Figures
\usepackage{subfigure}			% Sub figure system
\usepackage{hyperref}			% Links of references

\hypersetup{
  colorlinks   = true,	% Color links instead of ugly boxes
  urlcolor     = blue,	% Color of external hyperlinks
  linkcolor    = red,	% Color of internal links
  citecolor    = green	% Color of citations
}

\graphicspath{{./figures/}}

\begin{document}
% Title set-up
\title{Expanded evasion of the black hole no-hair theorem in dilatonic Einstein-Gauss-Bonnet theory}
\author{Bum-Hoon Lee}
\email{Electronic Address: bhl@sogang.ac.kr}
\affiliation{Department of Physics, Sogang University, Seoul 04107, Korea}
\affiliation{Center for Quantum Spacetime, Sogang University, Seoul 04107, Korea}
\author{Wonwoo Lee}
\email{Electronic Address: warrior@sogang.ac.kr}
\affiliation{Center for Quantum Spacetime, Sogang University, Seoul 04107, Korea}
\author{Daeho Ro}
\email{Electronic Address: daeho.ro@apctp.org}
\affiliation{Asia Pacific Center for Theoretical Physics, Pohang 37673, Korea}

\begin{abstract}
We study a hairy black hole solution in the dilatonic Einstein-Gauss-Bonnet theory of gravitation, in which the Gauss-Bonnet term is nonminimally coupled to the dilaton field. Hairy black holes with spherical symmetry seem to be easily constructed with a positive Gauss-Bonnet (GB) coefficient $\alpha$ within the coupling function, $f(\phi) = \alpha e^{\gamma \phi}$, in an asymptotically flat spacetime, i.e., no-hair theorem seems to be easily evaded in this theory. Therefore, it is natural to ask whether this construction can be expanded into the case with the negative coefficient $\alpha$. In this paper, we numerically present the dilaton black hole solutions with a negative $\alpha$, and we analyze the properties of GB term through the aspects of the black hole mass. We construct the new integral constraint allowing the existence of the hairy solutions with the negative $\alpha$. Through this procedure, we expand the evasion of the no-hair theorem for hairy black hole solutions.
\end{abstract}


\maketitle
%\tableofcontents

\section{Introduction} \label{sec:1}

The first astrophysical black hole is Cygnus A, which was later recognized as a black hole \cite{Reber}. The black hole is now a real thing, a most fascinating object, and worth exploring more deeply in the Universe. It was extensively investigated both observationally and theoretically. At the same time, various theories of gravitation, inspired by string theory or astrophysics were also developed. Based on these backgrounds, a variety of black hole solutions, such as a dilaton black hole \cite{Gibbons:1987ps, Horne:1992zy, Cai:1997ii} and Gauss-Bonnet (GB) black hole \cite{Boulware:1985wk, Cai:2001dz, Cai:2003gr}, have been studied. Furthermore, recent observations of a gravitational wave coming from the mergers of compact binary sources \cite{Abbott:2016nmj,Abbott:2017vtc} have opened new horizons in astrophysics as well as cosmology, in which it could be very interesting to test which theory of gravitation describes our Universe and the existence of the hairy black hole \cite{Pani:2009wy, TheLIGOScientific:2016src,Berti:2015itd}.

The existence of the hairy black hole solution with the GB term has been constructed and extensively studied over the past two decades \cite{Kanti:1995vq,Torii:1996yi,Kanti:1997br,Torii:1998gm,Guo:2008hf,Rogatko:2014maa,Herdeiro:2015waa,Myung:2018iyq}, in which the black hole has an exponentially decaying dilaton hair. Because of the motivation coming from string theory in general \cite{Boulware:1986dr,Callan:1988hs,Campbell:1991kz,Mignemi:1992pm}, the GB coefficient $\alpha$ is related to the Regge slope $\alpha'(=16\alpha)$ \cite{Torii:1996yi,Wiltshire:1988uq}, and it is always treated to be a positive constant on those works. From extensive studies, we noticed that there is the lower bound for a black hole mass, and that mass of a black hole increases when the dilaton coupling $\gamma$ increases \cite{Ahn:2014fwa,Gwak:2017fea}. We thought that the GB term seems to provide a repulsive property, which makes the formation of the dilaton black hole harder, and the lower bound increases as a result. What would happen if we change the sign of the GB coefficient? In this perspective, it is interesting to consider it as a kind of modified theory of gravitation, even though the motivation from string theory would not valid. For this reason, we more deeply investigate the dilaton black hole with the negative GB coefficient.

The no-hair theorem for black hole solutions was conjectured \cite{Ruffini:1971bza} to summarize the progress in black hole physics \cite{Oppenheimer:1939ue, Kerr:1963ud, Newman:1965my, Israel:1967wq, Israel:1967za, Carter:1971zc}, and developed \cite{Bekenstein:1972ny, Bekenstein:1995un} in Einstein-Maxwell theory, in which the solutions are associated with Gauss's law. In \cite{Bekenstein:1972ny}, the author used an integral constraint obtained from the equation of motion for the scalar field. Later, it was further developed into the novel no-hair theorem through the analysis of an energy-momentum tensor, especially the $T_r^{\ r}$ component \cite{Bekenstein:1995un}. If a black hole has the dilaton hair in the dilatonic Einstein-Gauss-Bonnet (DEGB) theory, the no-hair theorem should be avoided. Recently, we have seen that the no-hair theorem is bypassed for the black holes with a dilaton hair in DEGB theory \cite{Antoniou:2017acq, Antoniou:2017hxj}; by presenting both, the old no-hair theorem is easily evaded and the novel no-hair theorem is not applicable for DEGB theory. However, the GB coupling functions were positive definite in their analysis. For your interests, the no-hair theorem can also be evaded in the extended scalar-tensor-Gauss-Bonnet gravity \cite{Doneva:2017bvd,Doneva:2018rou} In this paper, we numerically present the dilaton black holes with a negative $\alpha$, and we analyze the properties of the GB term through the aspects of the lower bound for a black hole mass in more detail. The purpose of this paper is to provide the expanded evasion of the no-hair theorem for hairy black holes by constructing the new integral constraint to allow the existence of the dilaton black hole with arbitrary GB coefficients.

The paper is organized as follows: In Sec.\ \ref{sec:2}, we review and calculate the numerical setup. We analyze the energy momentum tensor and construct the new integral constraint. In Sec.\ \ref{sec:3}, we present a dilaton black hole solution with the negative GB coefficient , and we analyze the black hole properties with respect to the dilaton coupling and GB coefficient. In Sec.\ \ref{sec:4}, we summarize our results and discuss the role of the GB term with the difference between both cases.

\section{DEGB black hole} \label{sec:2}

%%%%%%%%%%%%%%%%%%%%%%%%%%%%%%%%%%%%%%%%%%%%%%%%%%%%%%%%%%%%
% Action, Einstein's equation and Scalar field equation
%%%%%%%%%%%%%%%%%%%%%%%%%%%%%%%%%%%%%%%%%%%%%%%%%%%%%%%%%%%%

Let us consider the action with the GB term:
\begin{equation}
S = \int d^4x \sqrt{-g} \left[\dfrac{R}{2} - \dfrac{1}{2} \nabla_\mu \phi \nabla^\mu \phi + f(\phi) R_{\rm GB}^2 \right] + S_{\rm b},
\end{equation}
where $g=\det g_{\mu\nu}$, the coupling function with the GB term is given by $f(\phi) = \alpha e^{\gamma \phi}$, and $\phi$ is a dilaton field. The scalar curvature of the spacetime is denoted by $R$, and the GB curvature term is given by $R_{\text{GB}}^2 = R^2 - 4R_{\mu\nu}R^{\mu\nu} + R_{\mu\nu\rho\sigma} R^{\mu\nu\rho\sigma}$. In this work, the boundary term $S_{\rm b}$ \cite{York:1972sj,Gibbons:1976ue,Myers:1987yn,Davis:2002gn} is not important, so it is abbreviated. The Einstein constant $\kappa = 8\pi G$ is set to unity for simplicity. The dilaton field equation is
\begin{equation}
\label{eq:scalar}
\dfrac{1}{\sqrt{-g}} \partial_\mu(\sqrt{-g} \partial^\mu \phi) + \dot{f}(\phi) R_{\text{GB}}^2 = 0,
\end{equation}
where the dot notation denotes the derivative with respect to $\phi$, and Einstein's equation is
\begin{equation}
R_{\mu\nu} - \dfrac{1}{2} g_{\mu\nu} R = T_{\mu\nu} = \partial_\mu \phi \partial_\nu \phi - \dfrac{1}{2} g_{\mu\nu} \partial_\rho \phi \partial^\rho \phi + T_{\mu\nu}^{\text{GB}},
\end{equation}
where $T_{\mu\nu}^{\rm GB}$ is the energy momentum tensor contributed from the GB term \cite{Nojiri:2005vv} as follows:
\begin{eqnarray}
\nonumber
T_{\mu\nu}^{\rm GB} &=& 4 (\nabla_\mu \nabla_\nu f(\phi)) R - 4 g_{\mu\nu} (\nabla^2 f(\phi)) R
\\ \nonumber
&& - 8 (\nabla_\rho \nabla_\mu f(\phi)) R_\nu{}^\rho - 8 (\nabla_\rho \nabla_\nu f(\phi)) R_\mu{}^\rho
\\ \nonumber
&& + 8 (\nabla^2 f(\phi)) R_{\mu\nu} + 8 g_{\mu\nu} (\nabla_\rho \nabla_\sigma f(\phi)) R^{\rho\sigma}
\\
&& - 8 (\nabla^\rho \nabla^\sigma f(\phi)) R_{\mu\rho\nu\sigma}.
\end{eqnarray}
The equation only has the derivative terms of $f(\phi)$ because the minimally coupled terms in four-dimensions are cancelled identically \cite{deWitt:1964}, and the contribution of the GB term with the equations of motion is coming from the nonminimally coupled terms only.

%%%%%%%%%%%%%%%%%%%%%%%%%%%%%%%%%%%%%%%%%%%%%%%%%%%%%%%%%%%%
% Metric and equations of motion
%%%%%%%%%%%%%%%%%%%%%%%%%%%%%%%%%%%%%%%%%%%%%%%%%%%%%%%%%%%%

Let us consider the spherically symmetric static metric in an asymptotically flat spacetime as follows:
\begin{equation}
ds^2 = - e^{X(r)} dt^2 + e^{Y(r)} dr^2 + r^2 (d\theta^2 + \sin^2 \theta d\varphi^2),
\end{equation}
where the metric functions $X$ and $Y$ depend only on $r$. Then, the dilaton field and Einstein's equations turn out to be \cite{Kanti:1995vq}
\begin{widetext}
\begin{subequations}
\begin{eqnarray}
\label{eq:df}
0 &=& \phi'' + \phi' \left( \dfrac{X'- Y'}{2} + \dfrac{2}{r} \right) - \dfrac{4\dot{f}}{r^2} \left( X' Y' e^{-Y} + (1-e^{-Y})\left(X'' + \dfrac{X'}{2}(X'-Y')\right)\right),
\\
\label{eq:tt}
0 &=& \dfrac{r}{2}\phi'^2 + \dfrac{1-e^Y}{r} - Y' \left(1+\dfrac{4\dot{f} \phi'}{r}(1-3e^{-Y})\right) + \dfrac{8\dot{f}}{r} \left(\phi'' + \dfrac{\ddot{f}}{\dot{f}}\phi'^2 \right) (1-e^{-Y}),
\\
\label{eq:rr}
0 &=& \dfrac{r}{2}\phi'^2 + \dfrac{1 - e^{Y}}{r} - X' \left(1 + \dfrac{4\dot{f} \phi'}{r}(1 - 3e^{-Y})\right),
\\ \nonumber
0 &=& X'' + \left(\dfrac{X'}{2} + \dfrac{1}{2} \right)(X' - Y') + \phi'^2 - \dfrac{8\dot{f}e^{-Y}}{r} \left(\phi' X'' + \left(\phi'' + \dfrac{\ddot{f}}{\dot{f}}\phi'^2 \right) X' + \dfrac{\phi'X'}{2}(X' - 3Y') \right),
\\
\label{eq:qq}
\end{eqnarray}
\end{subequations}
\end{widetext}
where the prime notation denotes the derivatives with respect to $r$. Equation \eqref{eq:rr} can be solved in terms of $Y$ as follows:
\begin{multline}
\label{eq:ey}
e^{Y(r)} = \dfrac{1}{4}\bigg(- r^2 \phi'^2 + 2r X' + 8 \dot{f} X' \phi' + 2 \\
 \pm \sqrt{(- r^2 \phi'^2 + 2r X' + 8 \dot{f} X' \phi' + 2)^2 - 192 \dot{f} X' \phi' } \bigg).
\end{multline}
We should take the positive sign from the above equation to be valid the near the horizon limit. In terms of the above equation, $Y$ and $Y'$ can be eliminated from the equations of motion. Thus, we use the Eqs.\ \eqref{eq:df} and \eqref{eq:qq} mainly for numerical calculation and the remaining one for constraint. For later use, it is better to calculate the GB term $R_{\rm GB}^2$, which is given by
\begin{equation}
R_{\rm GB}^2 =  \dfrac{2e^{-Y}}{r^2}[(1 - 3e^{-Y})X'Y'- (1 - e^{-Y})(X'^2+ 2X'')].
\end{equation}

%%%%%%%%%%%%%%%%%%%%%%%%%%%%%%%%%%%%%%%%%%%%%%%%%%%%%%%%%%%%
% Near horizon
%%%%%%%%%%%%%%%%%%%%%%%%%%%%%%%%%%%%%%%%%%%%%%%%%%%%%%%%%%%%

To perform the numerical computation, we impose the boundary conditions at the black hole horizon $r_h$ and asymptotically flat region $r \gg 1$. At the black hole horizon, the metric components should be zero such as $g_{tt}(r_h) = e^{X(r_h)} = 0$ and $g^{rr}(r_h) = e^{-Y(r_h)} = 0$. Then, the metric components and the dilaton field can be expanded in the near horizon limit by using the length parameter from the horizon, $\delta r = r - r_h$ as follows:
\begin{subequations}
\begin{eqnarray}
\label{eq:nhexpand1}
e^{X(r)} &=& 0 + x_1 \delta r + x_2 \delta r^2 + {\cal O}(\delta r^3),
\\
e^{-Y(r)} &=& 0 + y_1 \delta r + y_2 \delta r^2 + {\cal O}(\delta r^3),
\\
\phi(r) &=& \phi_h + \phi'_h \delta r + \phi''_h \delta r^2 + {\cal O}(\delta r^3).
\end{eqnarray}
\end{subequations}
The above equations provide the boundary conditions for $X$ and $\phi$, but the value $\phi_h$ is not determined yet. To do so, we expand the Eq.\ \eqref{eq:ey} in the near horizon limit:
\begin{multline}
\label{eq:eynr}
e^{Y(r)} = (r + 4\dot{f} \phi') X' \\
+ \dfrac{2r - r^3 \phi'^2 - 4\dot{f} \phi'(4 + r^2 \phi'^2)}{2(r + 4\dot{f} \phi')} + {\cal O} \left(\dfrac{1}{X'}\right).
\end{multline}
Now, we differentiate the equation with respect to $r$ and substitute the result from Eq.\ \eqref{eq:eynr} into Eqs.\ \eqref{eq:df} and \eqref{eq:qq}, as we discussed earlier. It eliminates $Y'$, and we only need to solve $X$ and $\phi$, not $Y$, when we solve the equation numerically. It is also possible to diagonalize the equations in terms of $X''$ and $\phi''$, but the result is not simple \cite{Kanti:1995vq,Ahn:2014fwa,Gwak:2017fea}. Finally, an expansion of the result in the near horizon limit gives the results
{\small
\begin{eqnarray}
\nonumber
X''(r) &=& - \dfrac{r^4 + 8r^3 \dot{f} \phi' + 16r^2 \dot{f}^2 \phi'^2 - 48 \dot{f}^2}{r^4 + 4r^3 \dot{f}\phi' - 96\dot{f}^2} X'^2 + {\cal O} \left(X'\right),
\\ \nonumber
\phi''(r) &=& - \dfrac{(r + 4\dot{f} \phi')(r^3\phi' + 4r^2 \dot{f}\phi'^2 + 12 \dot{f})}{r^4 + 4r^3 \dot{f}\phi' - 96\dot{f}^2} X'+ {\cal O} \left(1 \right).
\\ \label{eq:pp}
\end{eqnarray}}
One might notice that $\phi''$ will diverge at the horizon because of $X'$ diverges. Thus, the numerator should be zero, e.g.\ $r^3\phi' + 4r^2 \dot{f}\phi'^2 + 12 \dot{f} = 0$ not the other factor to be consistent with $e^{Y(r)}$, which indicates the value of $\phi'_h$ that is
\begin{equation}
\label{eq:phiph}
\phi'_h = - \dfrac{r_h^2 \pm \sqrt{r_h^4 - 192 \dot{f}_h^2}}{8r_h \dot{f}_h}.
\end{equation}
We take the negative sign to get the appropriate solution. It is easy to see that the positive sign will not give a solution that we want \cite{Kanti:1995vq}. To obtain the real value, there exists a restriction for $\phi_h$ from the square root term of $\phi'_h$, however, it is highly model dependent. Let us assume that $\phi_h$ is an arbitrary constant but it satisfies $r^4_h \geq 192 \dot{f}^2_h$. The substitution of $\phi'_h$ into Eqs.\ \eqref{eq:pp} at the horizon reduces the equations as follows:
\begin{equation}
\phi''(r) \approx 0, \qquad \text{and} \qquad X''(r) \approx -X'^2.
\end{equation}
Then, the derivative of metric function $X'$ is obtained,
\begin{equation}
X'(r) = \dfrac{1}{\delta r} + {\cal O} \left(1 \right),
\end{equation}
which recovers that which was originally assumed near the horizon limit, Eq.\ \eqref{eq:nhexpand1}. As a result, the GB term at the horizon is also obtained by
\begin{equation}
\label{eq:gbh}
R_{\rm GB}^2 \approx \dfrac{4 e^{-2Y}}{r^2}X'^2.
\end{equation}

%%%%%%%%%%%%%%%%%%%%%%%%%%%%%%%%%%%%%%%%%%%%%%%%%%%%%%%%%%%%
% Asymptotically flat region
%%%%%%%%%%%%%%%%%%%%%%%%%%%%%%%%%%%%%%%%%%%%%%%%%%%%%%%%%%%%

We also can expand the metric components and dilaton field in the asymptotically flat region in terms of the Arnowitt-Deser-Misner (ADM) mass $M$ \cite{Arnowitt:1959ah,Arnowitt:1962hi} and dilaton charge $D$ as follows:
\begin{subequations}
\begin{eqnarray}
e^{X(r)} &=& 1 - \dfrac{2M}{r} + {\cal O}(r^{-2}),
\\
e^{Y(r)} &=& 1 + \dfrac{2M}{r} + {\cal O}(r^{-2}),
\\
\phi(r) &=& \phi_\infty + \dfrac{D}{r} + {\cal O}(r^{-2}).
\end{eqnarray}
\end{subequations}
The GB term in the asymptotically flat region is then,
\begin{equation}
\label{eq:gba}
R_{\rm GB}^2 = \dfrac{48M^2}{r^6} + {\cal O}(r^{-7}).
\end{equation}
In order to focus on the numerical analysis, we will not show the relation between the parameters in the near horizon limit and asymptotically flat region in more detail (see the Refs.\ \cite{Antoniou:2017acq,Antoniou:2017hxj}). The ADM mass is represented as follows:
\begin{equation}
M = M(r_h) + M_{\rm hair},
\end{equation}
where the first term is the mass inside the horizon, $M(r_h) = r_h / 2$, and the second term is the mass of the dilaton hair. Once the metric is obtained numerically by the shooting method from the horizon, it is possible to obtain the mass and charge of the black hole by matching the behavior of the metric in the asymptotically flat region.

%%%%%%%%%%%%%%%%%%%%%%%%%%%%%%%%%%%%%%%%%%%%%%%%%%%%%%%%%%%%
% Novel no-hair theorem
%%%%%%%%%%%%%%%%%%%%%%%%%%%%%%%%%%%%%%%%%%%%%%%%%%%%%%%%%%%%

Since the dilaton black hole mass has the contribution coming from the existence of a scalar hair, it seems to evade the no-hair theorem. For this reason, we analyze whether or not there is a contradiction in the equations of motion with the energy-momentum tensor as in \cite{Antoniou:2017acq}. This is an important procedure to both compare and evade the novel no-hair theorem in \cite{Bekenstein:1995un}.

The $(tt)$ and  $(rr)$ components of the energy-momentum tensor are given by
\begin{subequations}
\begin{eqnarray}
\nonumber
T_t^{\ t} &=& - \dfrac{e^{-Y}}{2r^2} \bigg( r^2 \phi'^2 - 8 \dot{f} \phi' Y' (1 - 3 e^{-Y}) \\
&& + 16 (\dot{f} \phi'' + \ddot{f} \phi'^2) (1 - e^{-Y}) \bigg), \\
T_r^{\ r} &=& \dfrac{e^{-Y}}{2r^2} \left( r^2 \phi'^2 - 8\dot{f} \phi' X' (1 - 3e^{-Y}) \right), \\
\nonumber
(T_r^{\ r})' &=& \dfrac{e^{-Y}}{2r^2} \bigg( 2 r^2 \phi' \phi'' - r^2 \phi'^2 Y' - 8 \dot{f} \phi' X'' (1 - 3e^{-Y}) \\ \nonumber
&& - 8 (\dot{f} \phi'' + \ddot{f} \phi'^2 ) X' (1 - 3e^{-Y})  \\ \nonumber
&& + 8 \dot{f} \phi' X' Y' (1 - 6e^{-Y}) + \dfrac{16}{r} \dot{f} \phi' X' (1 - 3e^{-Y}) \bigg). \\
\end{eqnarray}
\end{subequations}
In the near horizon limit, $T_t^{\ t}$, $T_r^{\ r}$ and $(T_r^{\ r})'$ are reduced to
\begin{subequations}
\begin{eqnarray}
T_t^{\ t} &=& \dfrac{4 r^3 \dot{f} \phi' + 96 \dot{f}^2}{r^2(r^4 + 4 r^3 \dot{f} \phi' - 96 \dot{f}^2)} + {\cal O}(\delta r), \\
T_r^{\ r} &=& -\dfrac{4 \dot{f}\phi'}{r^2 (r + 4 \dot{f}\phi')} + {\cal O}(\delta r), \\
(T_r^{\ r})' &=& 0 \times X' + {\cal O}(1) + {\cal O}(\delta r).
\end{eqnarray}
\end{subequations}
Since $\dot{f}\phi'$ is negative definite from Eq.\ \eqref{eq:phiph} at the horizon, $T_r^{\ r}$ is positive definite in the near horizon limit. However, for $(T_r^{\ r})'$, the first order, which depends on $X'$, is identically zero, and so we should consider the next order. But, it is very hard to find and difficult to express, so we calculate the sign by putting whole horizon values into the second order, and we obtain the negative value. The sign is same as shown in \cite{Antoniou:2017acq}, even though we were not able to reproduce the results that they obtained. Similar to the case in the near horizon limit, those are given in the asymptotically flat region
\begin{multline}
- T_t^{\ t} = T_r^{\ r} = \dfrac{1}{2} \phi'^2 + {\cal O}(r^{-5}), \quad {\rm and} \\
(T_r^{\ r})' = \phi' \phi'' + {\cal O}(r^{-6}) = -\dfrac{2}{r} \phi'^2 + {\cal O}(r^{-6}),
\end{multline}
where we used the asymptotic relation $\phi'' = -(2/r) \phi' + {\cal O}(r^{-4})$ from Eq.\ \eqref{eq:df}. As a result, the tendency of $T_r^{\ r}$ and $(T_r^{\ r})'$ are summarized in Table \ref{tab:nohair}. We also plot the numerical results in Sec.\ \ref{sec:3.1} which correspond with our description. Thus, there is no contradiction in the equations of motion with the energy-momentum tensor, and we argue that the dilaton black hole evades the novel no-hair theorem even for the negative $\alpha$.
\begin{table}[htb]
\centering
\begin{tabular}{c|c|c}
&\\[-0.5em]
 & Near horizon region  & Asymptotically flat region \\[0.5em] \hline
&\\[-0.5em]
$T_t^{\ t}$ & $> 0$ & $< 0$ \\[0.7em] \hline
&\\[-0.5em]
$T_r^{\ r}$ & $> 0$ & $> 0$ \\[0.7em] \hline
&\\[-0.5em]
$(T_r^{\ r})'$ & $< 0$ & $< 0$ \\[0.7em]
\end{tabular}
\caption{Behavior summary of $T_r^{\ r}$ and $(T_r^{\ r})'$}
\label{tab:nohair}
\end{table}

%%%%%%%%%%%%%%%%%%%%%%%%%%%%%%%%%%%%%%%%%%%%%%%%%%%%%%%%%%%%
% Old no-hair theorem
%%%%%%%%%%%%%%%%%%%%%%%%%%%%%%%%%%%%%%%%%%%%%%%%%%%%%%%%%%%%

To be sure of our result, we also checked the old no-hair theorem, as shown in \cite{Antoniou:2017acq,Antoniou:2017hxj}, in which they developed the integral constraint equation with the positive definite coupling function $f(\phi)$ to show the evasion of the no-hair theorem. We construct the new integral constraint allowing the existence of the hairy solutions with arbitrary coupling functions. Starting with Eq.\ \eqref{eq:scalar}, it is possible to obtain the integral constraint,
\begin{multline}
\int f(\phi) \left( \nabla^2 \phi + \dot{f}(\phi) R_{\rm GB}^2 \right) \\
= - \int \dot{f}(\phi) \left( \phi'^2 - f(\phi) R_{\rm GB}^2 \right) = 0,
\end{multline}
where they used the integration in part only for the first term. The boundary term vanishes at the horizon and infinity due to the exponential factor of the metric and the derivative of the dilaton field, respectively \cite{Antoniou:2017acq}. Simply $\phi'^2$ is positive definite. The GB term is positive definite both on the horizon and in the asymptotically flat region. Thus, one can guess that the GB term is positive definite for all regions and is monotonically decreasing with respect to the radial length. Indeed, this really happens, and we will show the result in Sec.\ \ref{sec:3.1}. In order to avoid the no-hair theorem, the only condition for $f(\phi)$ is positive definite. In our study, we consider the negative $\alpha$, which makes the coupling function negative definite, in which the no-hair theorem seems to valid in this analysis. Therefore, we should find another way of treating the integral constraint and expand that, which covers all signs of definite cases of $f(\phi)$. As a result, we construct the integral constraint equation as follows:
\begin{multline}
\int e^{f(\phi)} \left( \nabla^2 \phi + \dot{f}(\phi) R_{\rm GB}^2 \right) \\
= - \int e^{f(\phi)} \dot{f}(\phi) \left( \phi'^2 - R_{\rm GB}^2 \right) = 0,
\end{multline}
where we also used the integration in part only for the first term. In the above equation, $\phi'^2$ and $R_{\rm GB}^2$ are positive definite and the coupling function $f(\phi)$ can be arbitrary. Thus, it is shown that the dilaton black hole solutions with arbitrary coupling functions evade the old no-hair theorem.

%%%%%%%%%%%%%%%%%%%%%%%%%%%%%%%%%%%%%%%%%%%%%%%%%%%%%%%%%%%%
% Boundary conditions
%%%%%%%%%%%%%%%%%%%%%%%%%%%%%%%%%%%%%%%%%%%%%%%%%%%%%%%%%%%%

In order to find the dilaton black hole solution, we used the Dormand-Prince method \cite{DORMAND198019}, which is one of the Runge-Kutta methods with specific parameters. Since the metric function diverges at the horizon, we start our calculation at $\delta r = \epsilon = 10^{-8}$, and we also set the infinity as $r_{\rm max} = 10^5$. Let us define the subscript $h$ and $\infty$ by means of the value at the initial point $r_h + \epsilon$ and the final point $r_{\rm max}$, respectively. Then, the initial conditions of the metric functions and field with the given coupling function $f(\phi) = \alpha e^{\gamma \phi}$ are
\begin{multline}
\label{eq:horizon}
X_h = \log(x_1 \epsilon), \quad X'_h = \dfrac{1}{\epsilon}, \quad \phi_h \leq \dfrac{1}{2\gamma} \log\left(\dfrac{r_h^4}{192\alpha^2 \gamma^2} \right), \\
\quad {\rm and} \quad \phi'_h = - \dfrac{r_h^2 - \sqrt{r_h^4 - 192 \dot{f}_h^2}}{8r_h \dot{f}_h}.
\end{multline}
One can see that the equations of motion are invariant under the shift of a dilaton field $\phi \rightarrow \phi + \phi_0$ for a constant $\phi_0$ with the rescaling of $r \rightarrow r e^{\gamma \phi_0 / 2}$ \cite{Kanti:1995vq}. Thus, we fix $r_h = 1$, vary $\phi_h$ to get a different black hole solution as a free parameter, and rescale the result. Until now, the parameter $x_1$ is arbitrary. On the other side, the boundary conditions at infinity are given by
\begin{equation}
X_\infty = 0, \quad X'_\infty = 0, \quad \phi = 0, \quad {\rm and} \quad \phi'_\infty = 0.
\end{equation}
To make $X_\infty = 0$, $x_1$ should be chosen properly, because the equations of motion depend only on $X'$ so the non-zero remaining constant $X_\infty$ can exist. In our calculation, we obtain the value by setting $x_1 = 1$, and we do the same procedure again with $\log x_1 = - X_\infty$, which is a way of fixing the parameter $x_1$. We also want to make the dilaton field vanish in the asymptotically flat region. For the dilaton field, the value $\phi_\infty$ also exists. The value can be absorbed by using the symmetry between $r$ and $\phi$ such as the rescaling of $r_h$ as $r_h \rightarrow r_h e^{-\gamma \phi_\infty / 2}$. Thus, the numerical calculation starts with same $r_h$ but can vary with the dilaton field value $\phi_h$, and the solutions form an one parameter family. Finally, we obtain $X$ and $\phi$ from the equations of motion and it is possible to obtain $Y$ by using Eq.\ \eqref{eq:ey}. The ADM mass $M$ and dilaton charge $D$ are obtained by fitting the equation in the asymptotically flat region.

\section{Results} \label{sec:3}

In this section, we present a hairy black hole solution in DEGB theory. We set the dilaton coupling function $f(\phi) = \alpha e^{\gamma \phi}$, where the GB coefficient $\alpha$ has the negative value, not the usual positive one. Since the DEGB theory has the rescaling invariance under the $r \rightarrow r / \sqrt{|\alpha|}$, we choose $\alpha = -1$ for all of our data. Furthermore, one can see that the theory is invariant under the changes of $\gamma \rightarrow -\gamma$ and $\phi \rightarrow -\phi$. Thus, we always choose the positive dilaton coupling $\gamma$, which is enough to obtain the solutions. Even in this unusual negative coefficient set up, we obtained the dilaton black hole solution and the different tendency of a minimum black hole mass depending on the $\gamma$.

\subsection{Dilaton black hole} \label{sec:3.1}

%%%%%%%%%%%%%%%%%%%%%%%%%%%%%%%%%%%%%%%%%%%%%%%%%%%%%%%%%%%%
% Solution example - figure 1,2
%%%%%%%%%%%%%%%%%%%%%%%%%%%%%%%%%%%%%%%%%%%%%%%%%%%%%%%%%%%%

This is an example of a hairy black hole solution with the negative $\alpha$. In order to get and present the dilaton black hole in this section, we set $\gamma = 1$ and $\phi_h = \log(r_h^4 / 192\alpha^2 \gamma^2) / 2\gamma$, which is the maximum value of the given range in the initial condition, Eq.\ \eqref{eq:horizon}.

\begin{figure}[t]
\centering
\subfigure[][\ $-g_{tt}(r)$ and $g_{rr}(r)$ vs. $r$]{\includegraphics[width=0.475\textwidth]{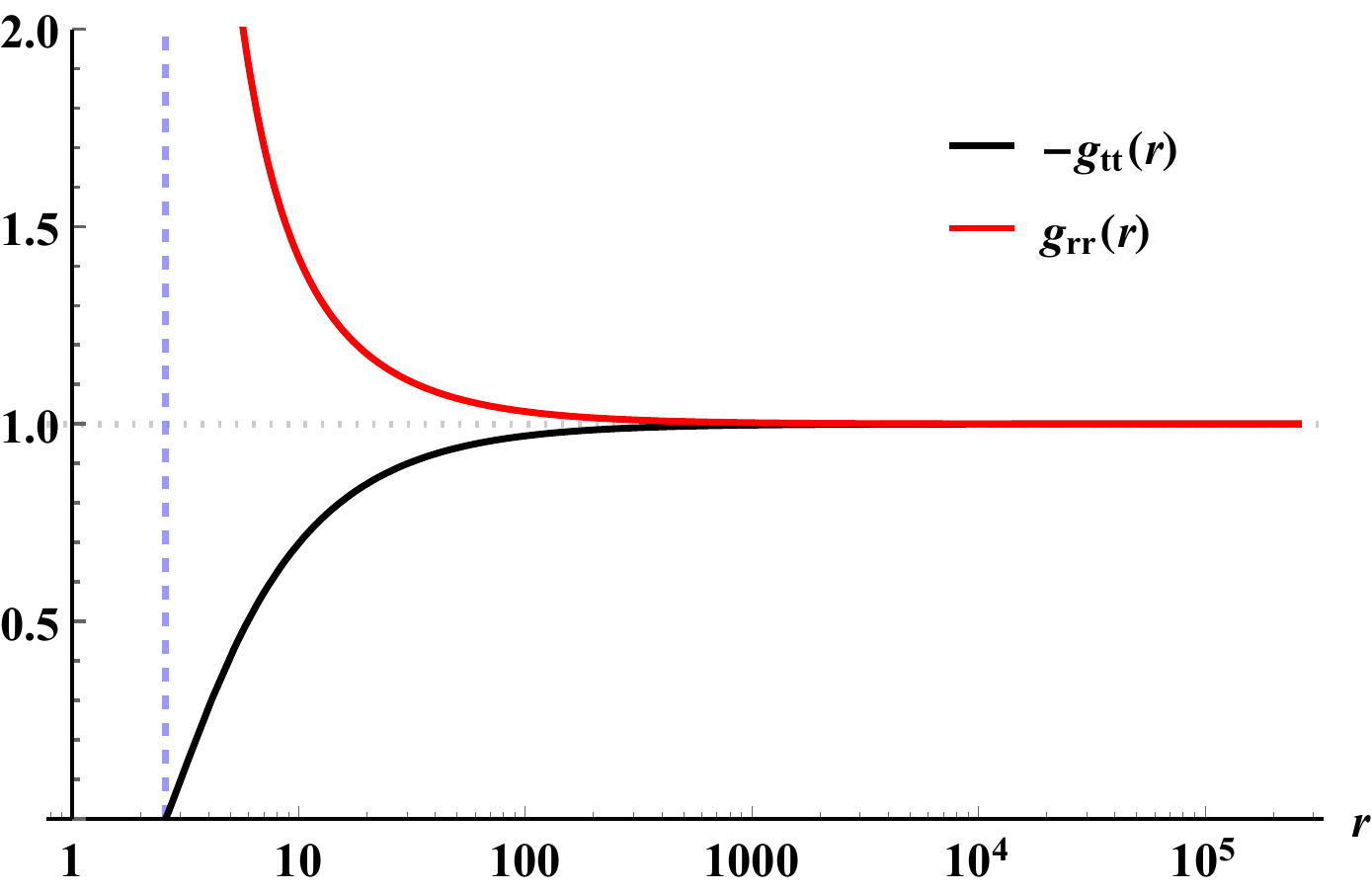}}
\subfigure[][\ $\phi(r)$ vs. $r$]{\includegraphics[width=0.475\textwidth]{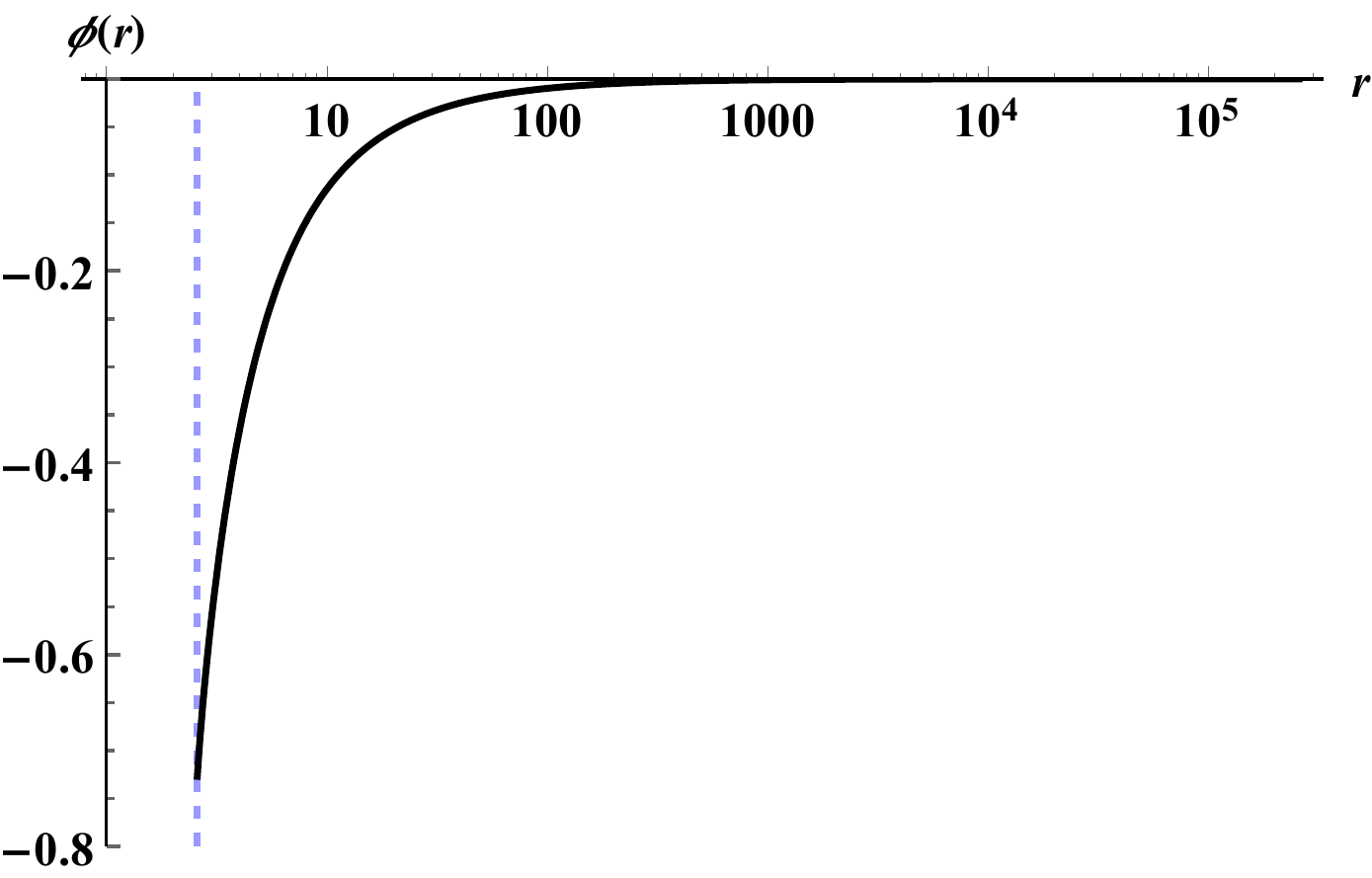}}
\caption{The metric components and the profile of the dilaton field for a black hole solution.}
\label{fig:metric}
\end{figure}

Figure \ref{fig:metric} represents the metric functions and the profile of the dilaton field for a black hole solution with respect to $r$. In Fig.\ \ref{fig:metric} (a), the black and red lines indicate the metric components $-g_{tt}(r)$ and $g_{rr}(r)$, which converge or diverge at the horizon, respectively. Both metric components converge to unity at infinity. In Fig.\ \ref{fig:metric} (b), the dilaton field $\phi(r)$ always has a negative value and the derivative of the dilaton field has a positive value. Both quantities also become zero at infinity. The blue dashed line in each figure indicates the value of the horizon radius, which is not unity. Originally, we set the horizon radius $r_h = 1$ but it is modified by the factor $e^{-\gamma \phi_\infty / 2}$, as we explained in the previous section.

\begin{figure}[t]
\centering
\subfigure[][\ $T_r^{\ r}(r)$ vs. $r$]{\includegraphics[width=0.475\textwidth]{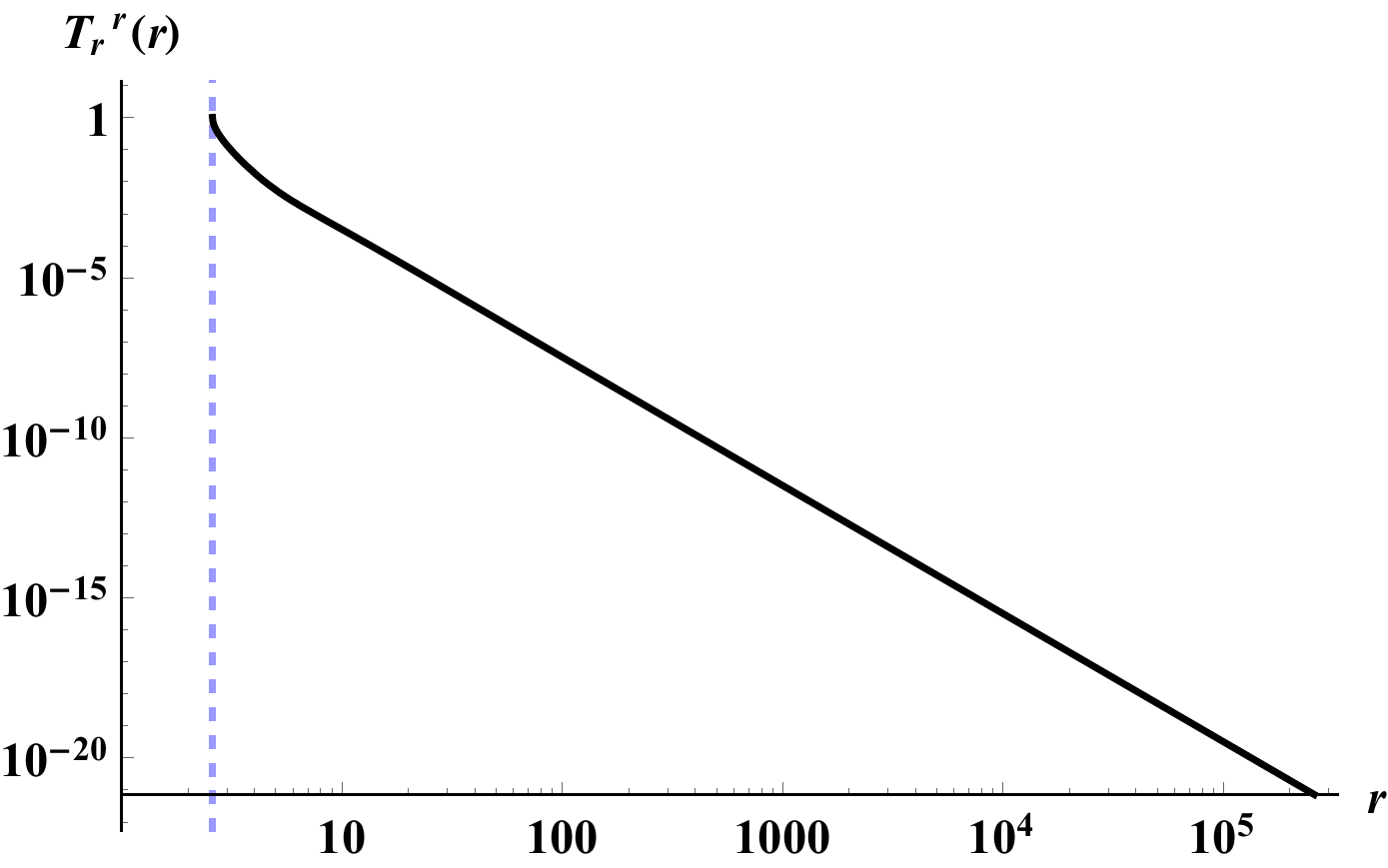}}
\subfigure[][\ $-(T_r^{\ r})'(r)$ vs. $r$]{\includegraphics[width=0.475\textwidth]{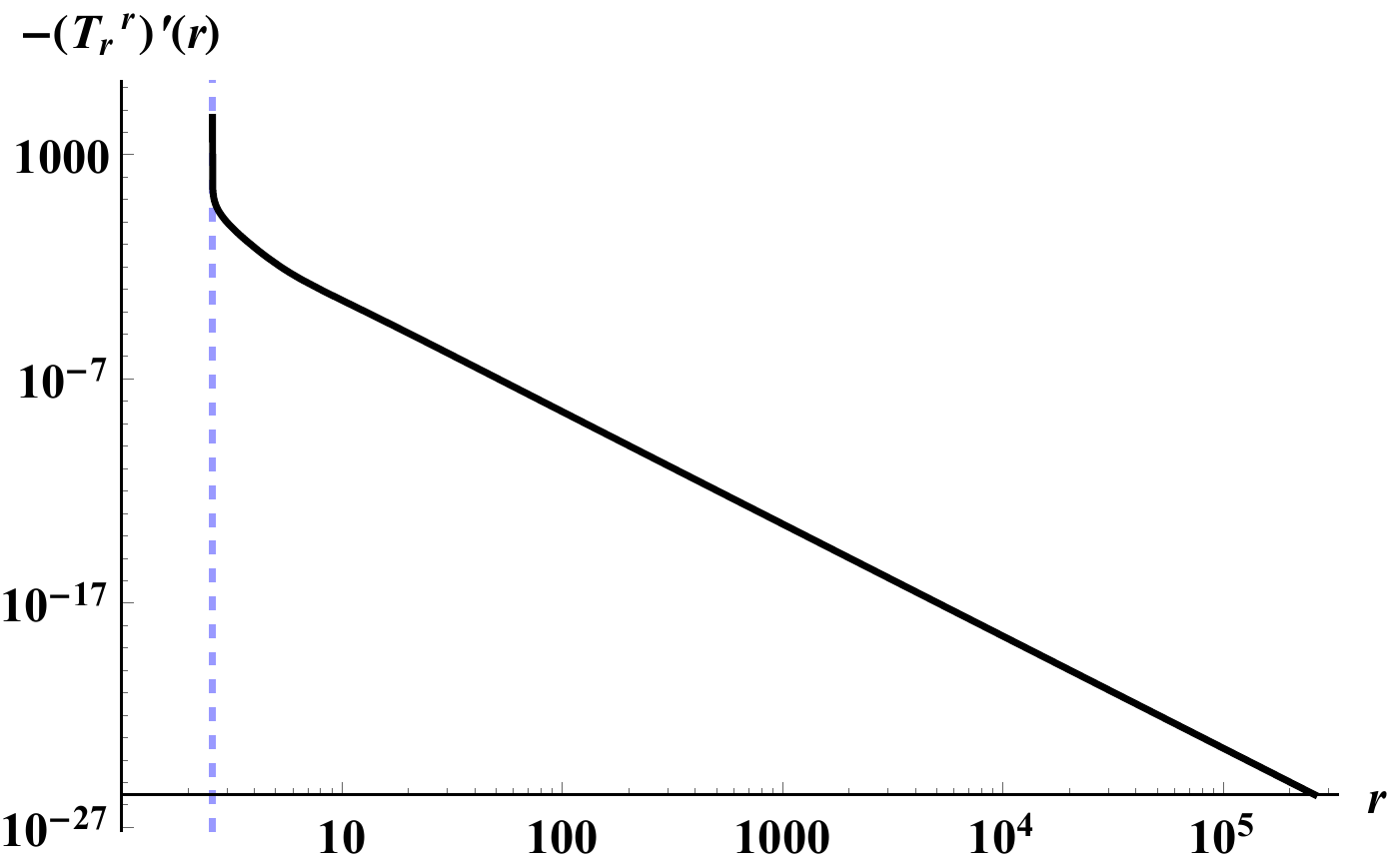}}
\caption{The $(rr)$ component of the energy momentum tensor and its derivative.}
\label{fig:emtensor}
\end{figure}

Figure \ref{fig:emtensor} shows the positivity of the $(rr)$ component of the energy-momentum tensor, $T_r^{\ r}(r)$, and the negative value of its derivative, $-(T_r^{\ r})'(r)$, with respect to $r$. Those quantities have positive values at the horizon and diminish when $r$ goes to infinity, but the signs never change. Therefore, there is no contradiction in the equations of motion with the energy-momentum tensor, which show that the novel no-hair theorem is not applicable and finally is not valid for the DEGB thoery, as we claimed before.

\begin{figure}[t]
\subfigure[][\ $R_{\rm GB}^2 (r)$ vs. $r$]{\includegraphics[width=0.475\textwidth]{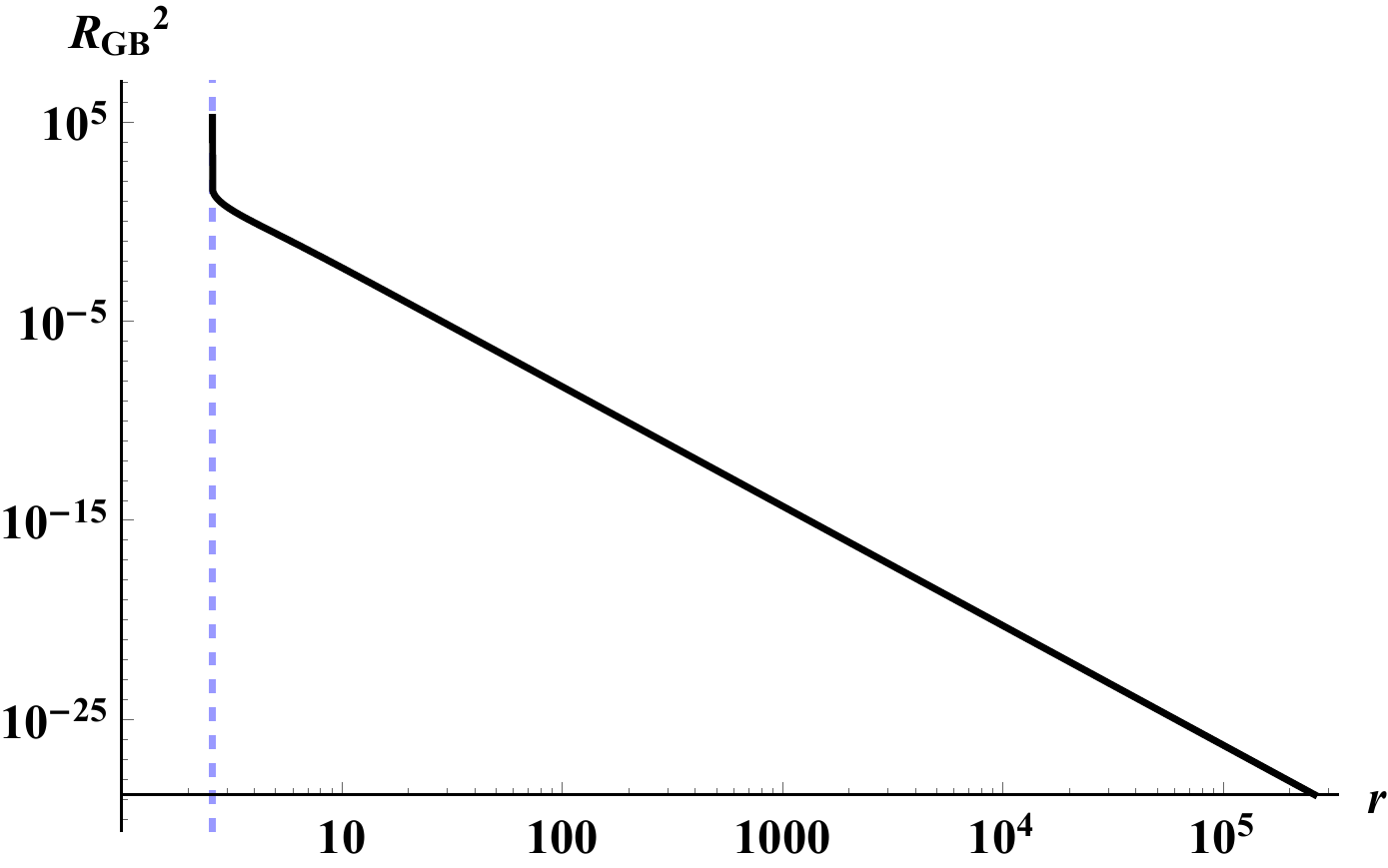}}
\subfigure[][\ -$T_t^{\ t}(r)$ and $T_r^{\ r}(r)$ vs. $r$]{\includegraphics[width=0.475\textwidth]{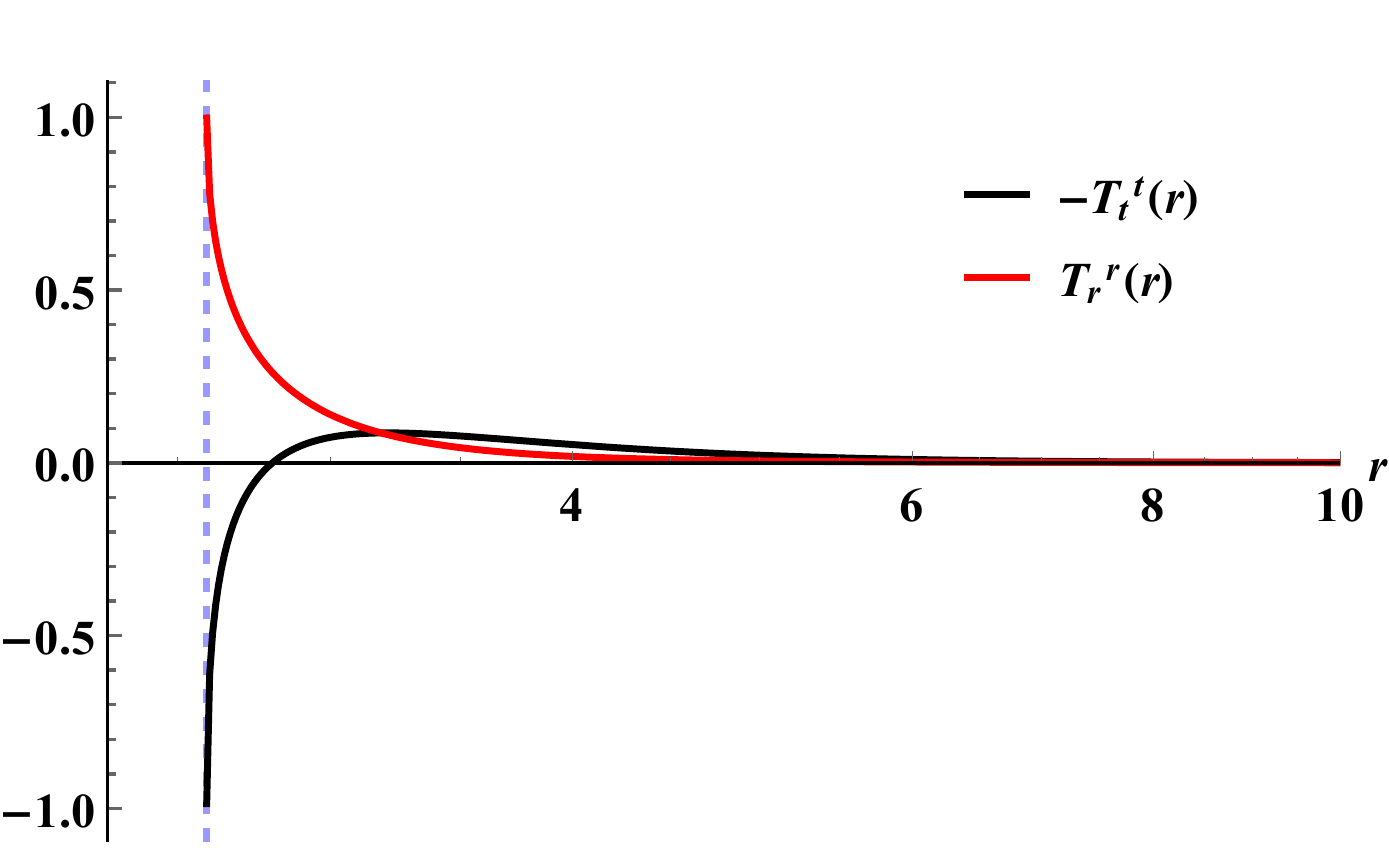}}
\caption{$R_{\rm GB}^2 (r)$ and the components of the energy momentum tensor.}
\label{fig:rgb}
\end{figure}

Figure \ref{fig:rgb} (a) illustrates the GB term $R_{\rm GB}^2 (r)$ with respect to $r$. The GB term is positive definite on all regions of $r$, and it shows the monotonically decreasing behavior when $r$ increases. The result corresponds well with our expectation about the old no-hair theorem, and the no-hair theorem is again evaded. In Fig.\ \ref{fig:rgb} (b), the black and red lines depict the $(tt)$ and $(rr)$ components of the energy-momentum tensor, $-T_t^{\ t}(r)$ and $T_r^{\ r}(r)$, respectively. The energy density $-T_t^{\ t}(r)$ has the negative value only for the near horizon region and the positive value for all the other regions of $r$. One of key assumptions in the novel no-hair theorem is related to the energy condition. The energy density is non-negative everywhere for any timelike observer. Thus, the existence of the negative value in some region shows the violation of the key assumptions satisfied in the novel no-hair theorem.

\subsection{Spectrum of dilaton black holes} \label{sec:3.2}

%%%%%%%%%%%%%%%%%%%%%%%%%%%%%%%%%%%%%%%%%%%%%%%%%%%%%%%%%%%%
% Solution spectrum with different \gamma - figure 3
%%%%%%%%%%%%%%%%%%%%%%%%%%%%%%%%%%%%%%%%%%%%%%%%%%%%%%%%%%%%

Now, we have the dilaton black hole solutions with the negative GB coefficient $\alpha$. In order to investigate the properties of the dilaton black holes with negative $\alpha$, we obtain the dilaton black hole solutions with same boundary conditions and compare them, but we just change the sign of $\alpha$ with respect to $\gamma$.

\begin{figure}[hbt]
\centering
\includegraphics[width=0.475\textwidth]{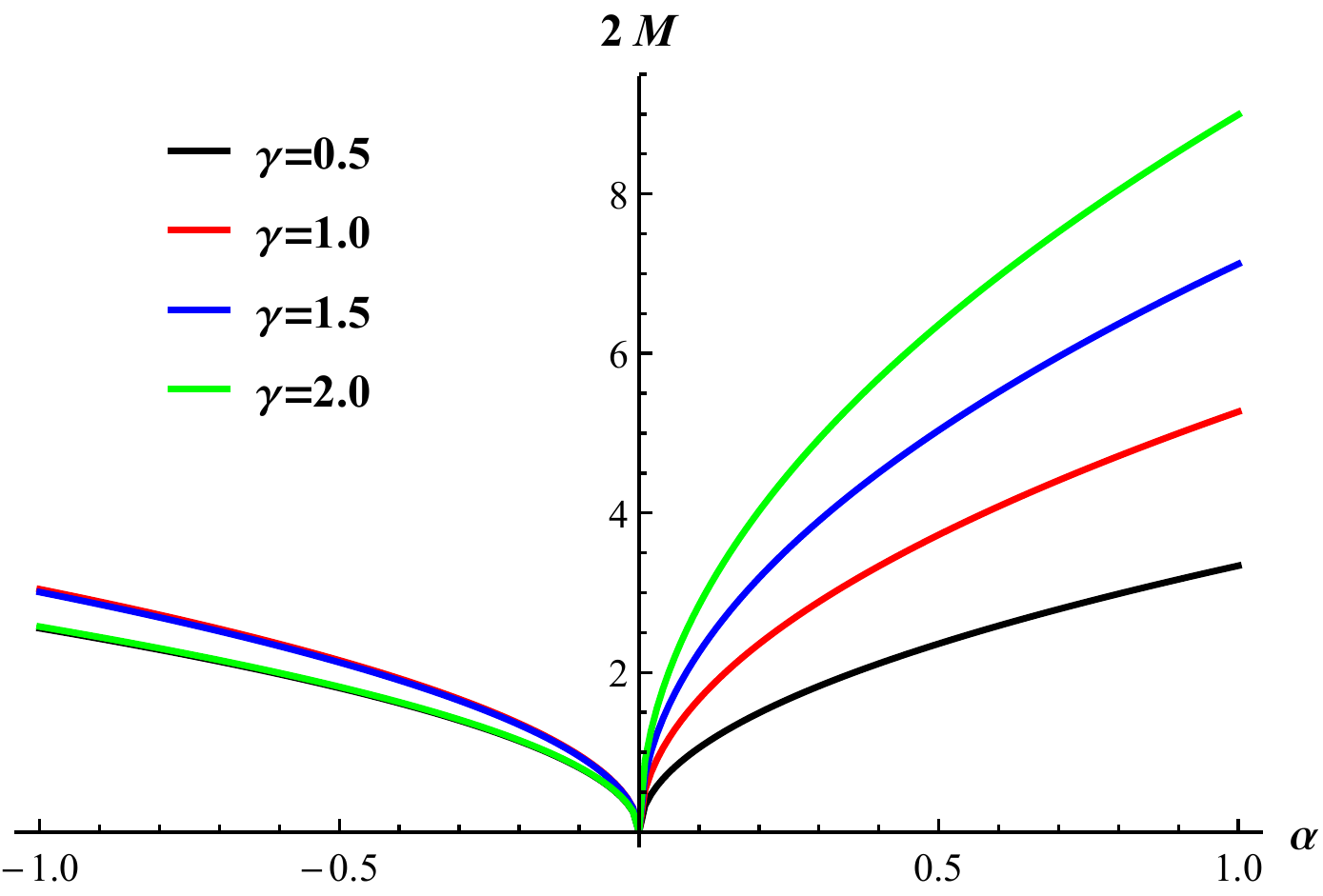}
\caption{The lower bound for the black hole mass vs. $\alpha$ with several $\gamma$ values.}
\label{fig:a_vs_m_wrt_gamma}
\end{figure}

Figure \ref{fig:a_vs_m_wrt_gamma} represents the lower bound for the dilaton black hole mass with respect to $\alpha$ for several selected values of $\gamma$. The $\phi_h$ is also chosen by the maximum value. It clearly shows that there exists the $\sqrt{|\alpha|}$ dependency of the black hole mass. The $\alpha$ dependency can be absorbed by the radial coordinate transformation, $r \rightarrow r / \sqrt{|\alpha|}$, as we discussed earlier. Therefore, we focused on the $\gamma$ dependency of the black hole mass for each sign of $\alpha$. The lower bound is increased when $\alpha$ has the positive value, but the lower bound is increased up to some specific $\gamma$ and decreased when $\alpha$ has the negative value as $\gamma$ is increased. Therefore, we expect that there exists some maximum $\gamma$ value, which restricts the dilaton black hole for the negative $\alpha$, and this is the most different behavior between the dilaton black holes with different signs of $\alpha$.

%%%%%%%%%%%%%%%%%%%%%%%%%%%%%%%%%%%%%%%%%%%%%%%%%%%%%%%%%%%%
% Main result - figure 4
%%%%%%%%%%%%%%%%%%%%%%%%%%%%%%%%%%%%%%%%%%%%%%%%%%%%%%%%%%%%

\begin{figure}[tb]
\centering
\subfigure[][\ The lower bound for the black hole mass vs. $\gamma$ with $\alpha = -1$.]{\includegraphics[width=0.475\textwidth]{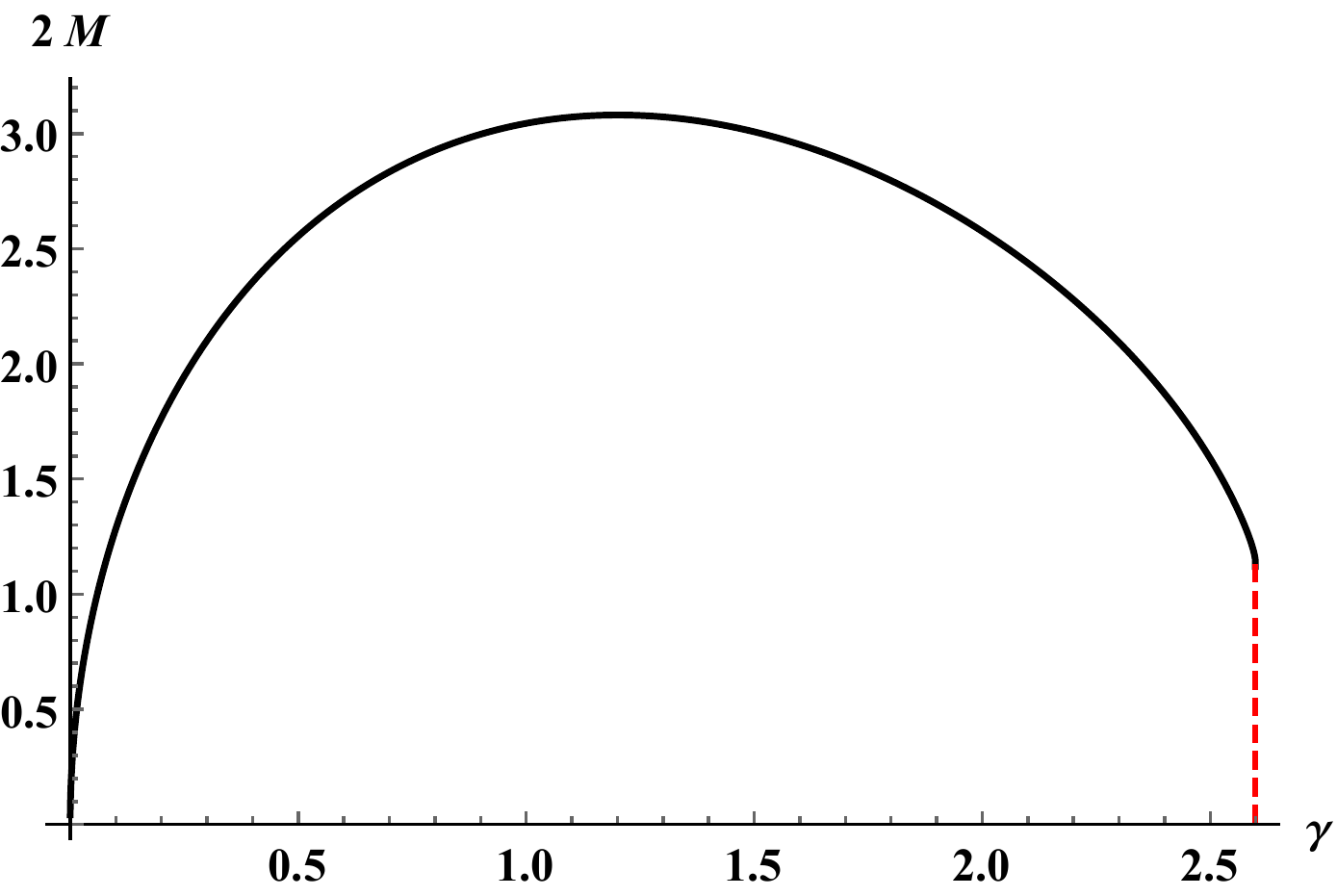}}
\subfigure[][\ The lower bound for the black hole mass vs. $\gamma$ with $\alpha = 1$.]{\includegraphics[width=0.475\textwidth]{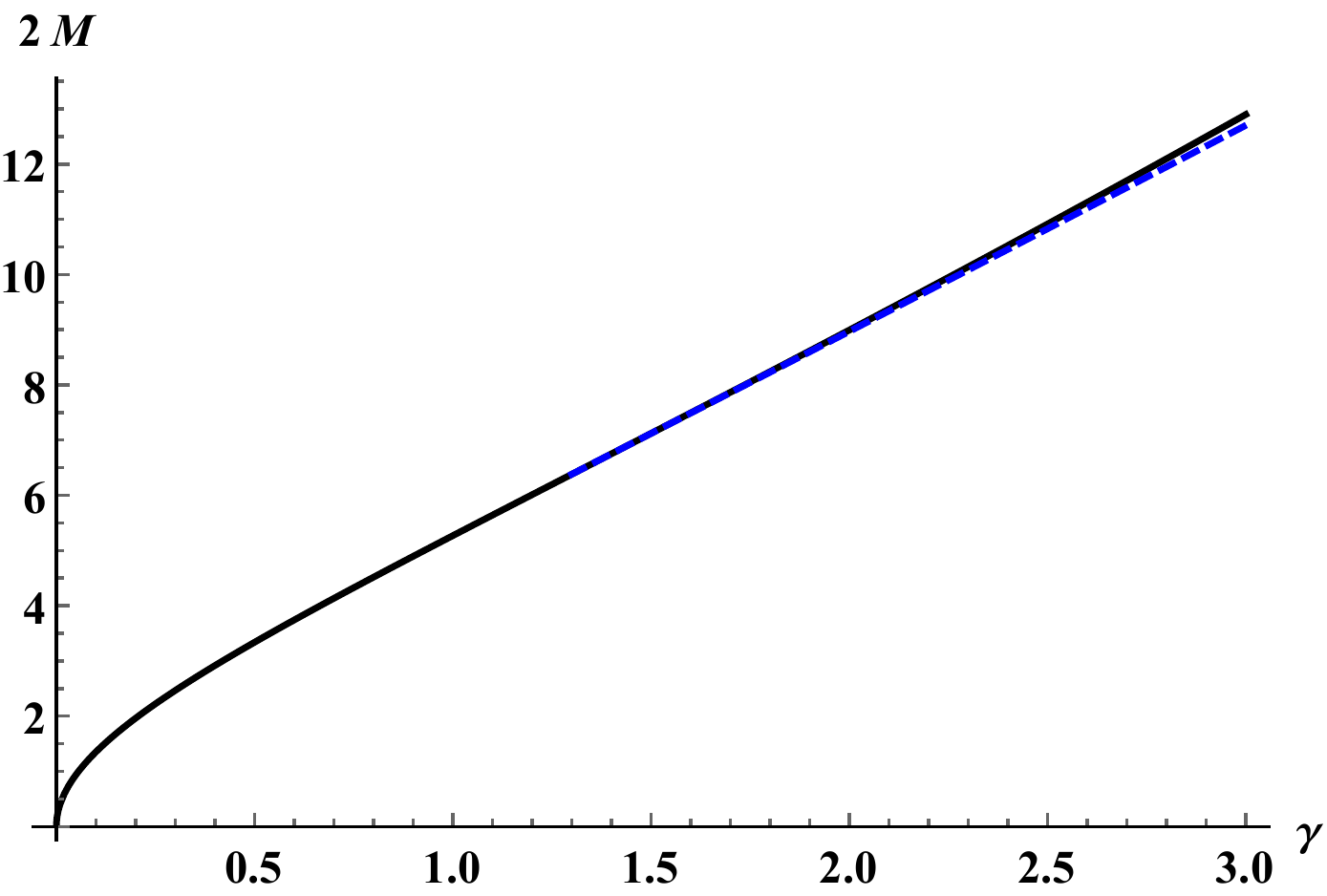}}
\caption{Several mass figures with respect to $\gamma$. The black line represent the lower bound for black holes with maximum value of $\phi_h$ which have the minimum black hole radius $r_h$. The red dashed line represents the maximized $\gamma$ to get the black hole solution with the negative $\alpha$. The blue dashed line represent the black holes having the minimum masses.}
\label{fig:minmass1}
\end{figure}

\begin{figure}[tb]
\subfigure[][\ The hairy mass vs. $\gamma$ with $\alpha = -1$.]{\includegraphics[width=0.475\textwidth]{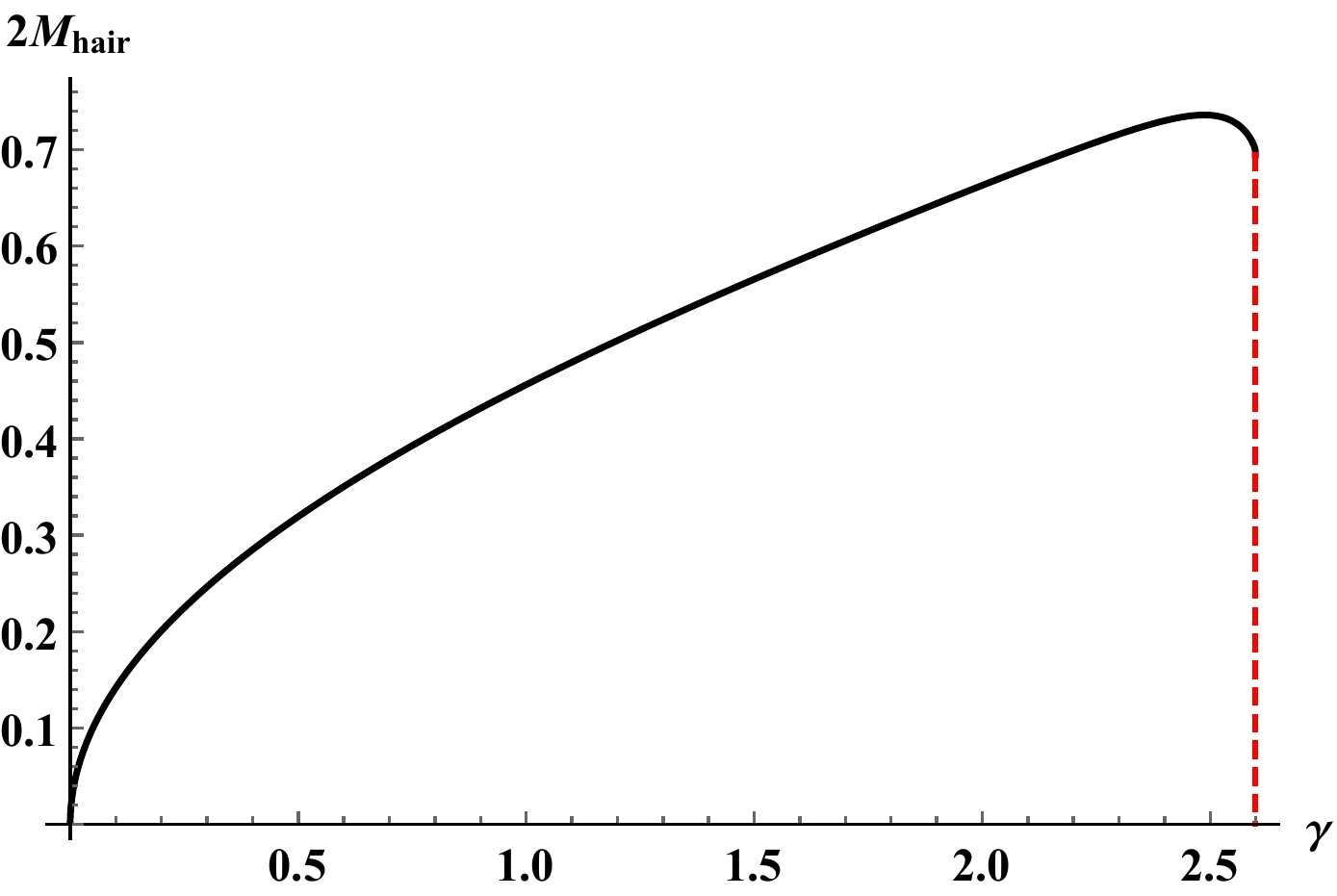}}
\subfigure[][\ The hairy mass vs. $\gamma$ with $\alpha = 1$.]{\includegraphics[width=0.475\textwidth]{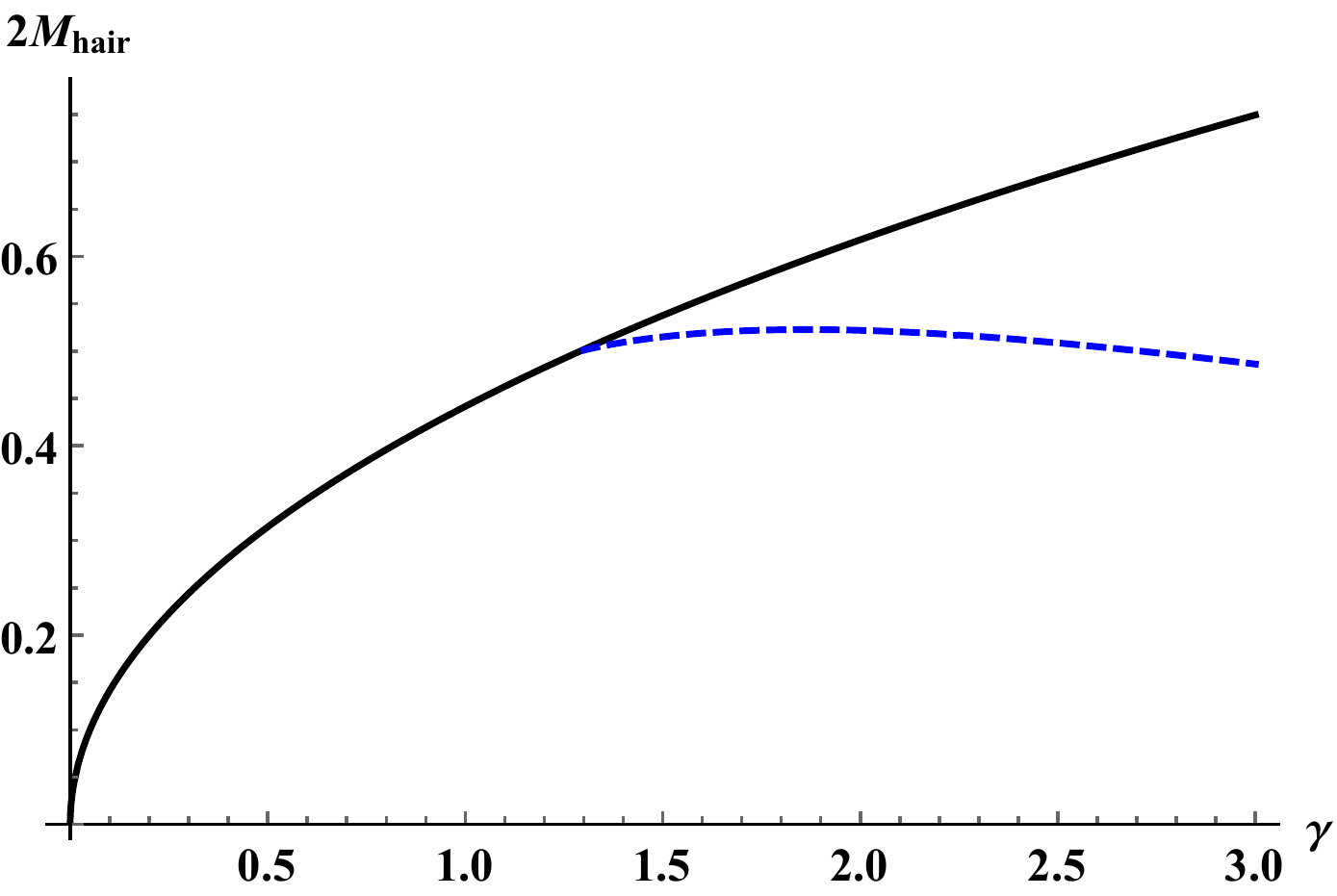}}
\caption{Several mass figures with respect to $\gamma$.}
\label{fig:minmass2}
\end{figure}

\begin{figure}[tb]
\subfigure[][\ The hairy mass ratio vs. $\gamma$ with $\alpha = -1$.]{\includegraphics[width=0.475\textwidth]{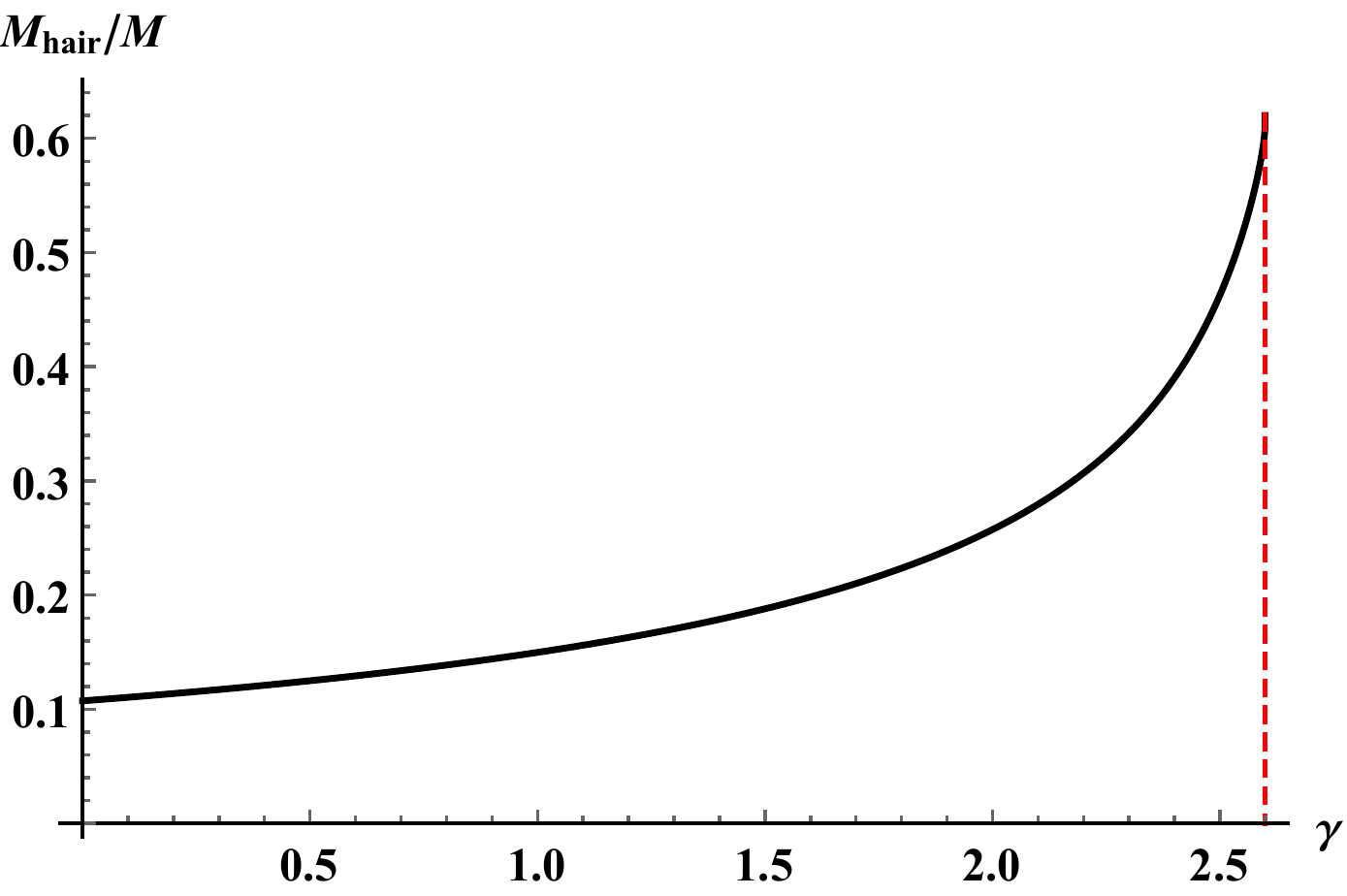}}
\subfigure[][\ The hairy mass ratio vs. $\gamma$ with $\alpha = 1$.]{\includegraphics[width=0.475\textwidth]{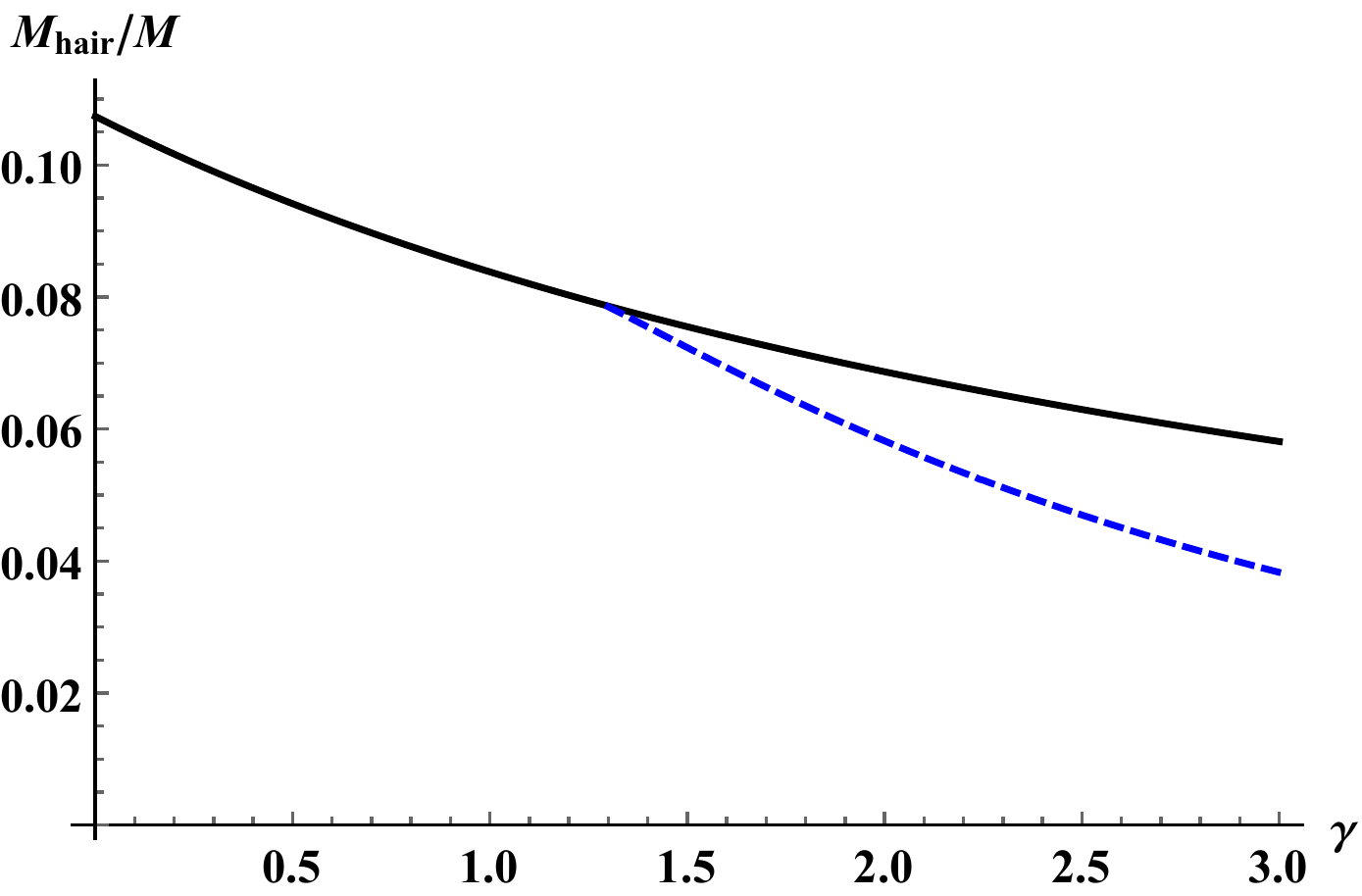}}
\caption{Several mass figures with respect to $\gamma$.}
\label{fig:minmass3}
\end{figure}

Figures \ref{fig:minmass1}, \ref{fig:minmass2}, and \ref{fig:minmass3} illustrate the mass of dilaton black holes with respect to $\gamma$ for each of the $\alpha$ signs. The black line denotes the black hole mass with the maximum value of $\phi_h$. The red dashed line indicates the maximum $\gamma$, which can have the dilaton black hole when $\alpha$ is negative. The blue dashed line represents the dilaton black hole, which has the minimum mass under the variations of $\phi_h$ when $\alpha$ is positive. It is already a well-known result that the maximum $\phi_h$ does not always give the minimum mass of the dilaton black hole \cite{Kanti:1997br,Torii:1998gm,Ahn:2014fwa,Gwak:2017fea}. The black and blue lines show the dilaton black hole with maximum $\phi_h$ or minimum mass. By changing $\phi_h$, we obtain the dilaton black holes with a higher mass than the ones represented by black or blue lines. Therefore, there also exists the dilaton black holes over the black or blue lines but not under the lines.

Figures \ref{fig:minmass1} (a) and \ref{fig:minmass1} (b) show the dilaton black hole mass increases and decreases when $\alpha$ is negative but it keep increasing even for the black holes with minimum masses when $\alpha$ is positive by increasing $\gamma$, as we have shown in Fig.\ \ref{fig:a_vs_m_wrt_gamma}. When $\alpha$ is negative, we cannot obtain the dilaton black hole solution with the $\gamma$ value which exceeds the red dashed line. In order to get the contribution coming from the GB term in more detail, we need to investigate the large $\gamma$ region, because the GB term is  exponentially dependent on the $\gamma$. In this region, it seems that the GB term decreases its own repulsive property and relatively easily assists in making the dilaton black hole. As a result, we argue that the minimum mass of the dilaton black hole decreases in the large $\gamma$ region. However, when $\alpha$ is positive, the GB term appears to demonstrate a dispersive behavior, and it disturbs the creation of the the dilaton black hole. Thus, we also argue that the minimum mass of the dilaton black hole increases depending on $\gamma$.

Figures \ref{fig:minmass2} (a) and \ref{fig:minmass2} (b) show the hairy mass of the black hole, and Figs.\ \ref{fig:minmass3} (a) and \ref{fig:minmass3} (b) show the ratio between the hairy mass and the total mass with respect to $\gamma$. The hairy mass increases in general, except the near maximum $\gamma$ region of the case with the negative $\alpha$ and the minimum mass black hole of the case with the positive $\alpha$. However, the ratio always increases or decreases when $\alpha$ is negative or positive, respectively. Especially for the case with the negative $\alpha$, the ratio increases even higher than $> 0.5$ near the maximum $\gamma$; the red line and the behavior are really strange. We wonder whether or not the value will keep growing until it reaches unity, which means that the black hole horizon would disappear and have the mass only as the dilaton hair. This does not happen when $\alpha$ is positive. The maximum and restricted value of $\gamma$ seems to be motivated from this reason. However, we cannot do an exact numerical calculation on the limit of maximum $\gamma$ due to the difficulties of error control, thus our argument remains as a reasonable but open question.

\section{Conclusion} \label{sec:4}

We have investigated the hairy black hole solutions in DEGB theory, in particular with the negative GB coefficient $\alpha$. In Refs.\ \cite{Antoniou:2017acq, Antoniou:2017hxj}, the authors showed that the no-hair theorems are easily evaded by the hairy black hole solutions in DEGB theory. They considered the black hole solutions with only the positive $\alpha$ and many scalar couplings, and they constructed the integral constraint in the theory.
In this paper, we tried to expand the description about the dilaton black hole into negative GB coefficients by changing the sign of $\alpha$. We constructed the new integral constraint equation allowing the existence of the hairy black hole solution with the arbitrary signature of $\alpha$. Through this procedure, we have expanded the evasion of the no-hair theorem for hairy black hole solutions.

As a consequence of our analysis, we have numerically obtained the dilaton black hole solutions with the negative $\alpha$. The dilaton black holes have more hair, in general, than the case for the positive $\alpha$. We restricted our calculation to the dilaton black holes that have the maximum values of dilaton fields at the horizon $\phi_h$ or have minimum masses. It is enough to investigate the properties of the dilaton black hole. The minimum mass of a dilaton black hole with a positive $\alpha$ is obtained from the maximum $\phi_h$ in the small dilaton coupling $\gamma$ region. When $\gamma$ is increased, the cases of dilaton black holes with the minimum mass and the maximum $\phi_h$ are divided. The mass of a dilaton black hole increases in both cases. We think that the GB term seems to provide a repulsive effect, and it disturbs the formation the dilaton black hole. However, those two cases are not divided with the negative $\alpha$ and the minimum mass decreases for large $\gamma$. Since the minimum mass decreases, it seems that the GB term decreases its own repulsive property, and the black hole forms relatively easy.

Furthermore, there exists a maximum value $\gamma$ that limits the existence of the dilaton black hole solution. Until the maximum $\gamma$, the minimum mass decreases but the hairy mass increases and decreases again. Interestingly, the hairy mass ratio of the total mass always increases. The results give us an expectation that the dilaton black hole solution with the maximum $\gamma$ would have no horizon, and the mass of the black hole is composed by the dilaton field only. Even though this expectation has yet ti be fulfilled, due to the difficulties of the numerical calculation, it is worthwhile to investigate the properties and implications of such behaviors for large $\gamma$ in more detail, and we postpone further analysis for future work.

It is important to investigate the stability for the maximally symmetric background as well as the hairy black hole solutions. For the maximally symmetric background, there is nonperturbative instability due to the tunneling for a nucleation of a vacuum bubble when the initial vacuum state is in the metastable vacuum and the true vacuum state exists \cite{Coleman:1977py, Coleman:1980aw, Lee:2008hz, Charmousis:2008ce}. For black hole solutions, there is no instability in Einstein theory when making use of the Regge-Wheeler prescription \cite{Regge:1957td}, while the stability issue is nontrivial in DEGB theory. There exists the positive lower-bound for the black hole mass in that theory. There are two black holes for a given mass above the lower-bound in which the smaller one is unstable and the larger one is stable under perturbations \cite{Torii:1996yi,Torii:1998gm}. The equations governing the perturbations of the metric are decoupled from the equation governing the perturbation of the scalar field \cite{Doneva:2017bvd}. Recently, it has been reported that the black hole without hair becomes unstable against scalar perturbations, and a new black hole solution with scalar hair bifurcates from the one without hair in DEGB \cite{Doneva:2017bvd, Silva:2017uqg} and Einstein-Maxwell scalar theory \cite{Herdeiro:2018wub}.

The higher-dimensional black hole in EGB theory was first discovered in Ref.\ \cite{Boulware:1985wk} and in Einstein-Maxwell-Gauss-Bonnet theory \cite{Wiltshire:1985us}, in which the lower bound for a black hole mass is proportional to $\alpha$, not $\sqrt{\alpha}$. However, there is no bound for the mass in six and higher dimensions \cite{Guo:2008hf, Sahabandu:2005ma}. It will be interesting to investigate our results in comparison with those in higher-dimensional DEGB theory or EGB theory.

The stability of the maximally symmetric vacuum in DEGB theory is not complicated. Let us consider the maximally symmetric vacuum spacetime, flat Minkowski. One can consider linear perturbations around this background. Then the quadratic curvature terms in the field equations should not contribute to the perturbation equations as in the higher-dimensional EGB theory \cite{Boulware:1985wk}. Therefore, the perturbation equations are the same as those in general relativity with a massless scalar field, and hence, the flat background is stable against linear perturbations for any value of $\alpha$ and $\gamma$. This can be confirmed for spherically symmetric linear perturbations \cite{Torii:1998gm}. When the background is flat Minkowski, the perturbation equation for the dilaton perturbation reduces to the Klein-Gordon equation in a two-dimensional flat spacetime that does not contain $\alpha$ and $\gamma$ as well as any higher-curvature terms. Thus, the flat background is stable against these spherical perturbations \footnote{We would like to thank an anonymous referee for useful comments and pointing this out.}.

We postpone any possible application for the evasion of the no-hair theorem and the applications for black hole solutions including the stability issue in higher-dimensional DEGB theory for our future work.

\acknowledgments
B.-H. Lee(NRF-2018R1D1A1B07048657), W. Lee (NRF-2016R1D1A1B01010234) and D. Ro (NRF-2017R1D1A1B03029430) were supported by Basic Science Research Program through the National Research Foundation of Korea funded by the Ministry of Education. D. Ro was also supported by the Korea Ministry of Education, Science and Technology, Gyeongsangbuk-Do and Pohang City. We would like to thank Hyeong-Chan Kim and Youngone Lee for their
hospitality during our visit to Korea National University of Transportation, Seoktae Koh to Jeju National University,
and Yun Soo Myung, Hyung Won Lee, Kyoung Yee Kim, Jeongcho Kim and De-Cheng Zou to Inje University.

\bibliographystyle{apsrev4-1}
\bibliography{ref}
\end{document}